\documentclass[journal,10pt]{IEEEtran}
\pdfoutput=1
\usepackage{comment}
\usepackage{lipsum}
\usepackage{mathtools}
\usepackage{graphicx}
\usepackage{bmpsize}
\usepackage{subfig}

\usepackage{float}
\usepackage{cite}
\usepackage{epstopdf}
\usepackage{multirow}
\usepackage{amssymb}
\usepackage{amsmath}
\usepackage{mathtools, cases}
\usepackage{chngcntr}
\usepackage{amsmath,etoolbox}
\usepackage{algorithmic}
\usepackage{amssymb}
\usepackage[ruled]{algorithm2e}
\usepackage[hyphens]{url}
\usepackage[
bookmarksopen,
bookmarksdepth=2,
breaklinks=true
]{hyperref}
\usepackage[mathscr]{euscript}

\usepackage[font=small,skip=3pt]{caption}
\usepackage{cleveref}
\usepackage{color}
\usepackage{scalerel,stackengine}

\begin{document}

\title{IEEE 802.11bd \& 5G NR V2X: Evolution of Radio Access Technologies for V2X Communications}
\author{Gaurang Naik, Biplav Choudhury, Jung-Min~(Jerry)~Park\thanks{The authors are with the Bradley Department of Electrical and Computer Engineering, Virginia Tech. (Email:\{gaurang, biplavc, jungmin\}@vt.edu)}\vspace{-8mm}}

\graphicspath{{images/}}
{}

\maketitle

\begin{abstract}
With rising interest in autonomous vehicles, developing radio access technologies (RATs) that enable reliable and low latency vehicular communications has become of paramount importance. Dedicated Short Range Communications (DSRC) and Cellular V2X (C-V2X) are two present-day technologies that are capable of supporting day-1 vehicular applications. However, these RATs fall short of supporting communication requirements of many advanced vehicular applications, which are believed to be critical in enabling fully autonomous vehicles. Both DSRC and C-V2X are undergoing extensive enhancements in order to support advanced vehicular applications that are characterized by high reliability, low latency and high throughput requirements. These RAT evolutions---IEEE 802.11bd for DSRC and NR V2X for C-V2X---can supplement today's vehicular sensors in enabling autonomous driving. In this paper, we briefly describe the two present-day vehicular RATs. In doing so, we highlight their inability to guarantee quality of service requirements of many advanced vehicular applications. We then look at the two RAT evolutions, i.e., IEEE 802.11bd and NR V2X and outline their objectives, describe their salient features and provide an in-depth description of key mechanisms that enable these features. While both, IEEE 802.11bd and NR V2X, are in their initial stages of development, we shed light on their preliminary performance projections and compare and contrast the two evolutionary RATs with their respective predecessors. 
\end{abstract}

\begin{IEEEkeywords}
V2X, DSRC, C-V2X, IEEE 802.11bd, NR V2X
\end{IEEEkeywords}

\IEEEpeerreviewmaketitle
\section{Introduction}


Vehicle-to-Everything (V2X) communications has the potential to significantly bring down the number of vehicle crashes, thereby reducing the number of associated fatalities~\cite{nhtsa-factsheet}. However, the benefits of V2X are not limited to safety applications alone. V2X-capable vehicles can assist in better traffic management leading to greener vehicles and lower fuel costs~\cite{wang2017overview}. Intelligent Transportation Systems (ITS) constitute such vehicular safety and non-safety applications. Today, the two key radio access technologies (RATs) that enable V2X communications are Dedicated Short Range Communications (DSRC) and Cellular-V2X (C-V2X). DSRC is designed to primarily operate in the 5.9 GHz band, which has been earmarked in many countries for ITS applications. On the other hand, C-V2X can operate in the 5.9 GHz band as well as in the cellular operators' licensed carrier~\cite{wang2017overview}. 

DSRC relies on the IEEE 802.11p standard for its physical (PHY) and medium access control (MAC) layers. DSRC uses a MAC protocol that is simple, well-characterized and capable of distributed operations. However, the adoption of DSRC in vehicles has been delayed due to its poor scalability and communication challenges imposed by high-mobility environments. Meanwhile, the 3rd Generation Partnership Project (3GPP) has developed C-V2X---a Long Term Evolution (LTE) based RAT---that can enable C-V2X capable vehicles to operate in a distributed manner in the absence of cellular infrastructure, while leveraging the infrastructure for efficient resource allocations when vehicles operate within coverage.   

Existing literature shows that C-V2X offers performance advantages over DSRC in terms of its additional link budget, higher resilience to interference and better non line-of-sight (NLOS) capabilities~\cite{fcc_adj}. Further, studies indicate that both DSRC and C-V2X can reliably support safety applications that demand an end-to-end latency of around $100$~milliseconds (msec) as long as the vehicular density is not very high~\cite{molina2017lte}. However, as the quality of service (QoS) requirements of V2X use-cases become more stringent, which is the case in many advanced V2X applications~\cite{tr22886}, the two current V2X RATs fall short of providing the desired performance.

In order to diminish the performance gap between DSRC and C-V2X and to support additional modes of operations and increase the offered throughput, a new Study Group called the IEEE 802.11 Next Generation V2X was formed in March 2018~\cite{IEEEP80223:online}. This resulted in the formation of IEEE Task Group 802.11bd (TGbd) in Jan. 2019. On the other hand, 3GPP is working toward the development of New Radio (NR) V2X for its Rel. 16, building atop of 5G NR that was standardized in 3GPP Rel.15. NR V2X is expected to support advanced V2X applications that require much more stringent QoS guarantees compared to applications that can be supported by C-V2X~\cite{tr22886}. Some of these use-cases require the end-to-end latency to be as low as $3$~msec with a reliability of $99.999\%$! Coupled with the existing challenges offered by high-mobility environments, these additional constraints make the design of 802.11bd and NR V2X extremely challenging.

In terms of their design objectives, 802.11bd and NR V2X have certain similarities. For example, both evolutionary RATs are being designed to improve the reliability of offered services, lower the end-to-end latency and support applications that require high throughput. However, their design methodologies significantly differ. TGbd requires the new standard, i.e., 802.11bd to be backward compatible with 802.11p. This implies that 802.11bd and 802.11p devices must be able to communicate with each other while operating on the same channel. On the other hand, 3GPP does not impose a similar constraint on NR V2X. Vehicles equipped with NR V2X can still communicate with C-V2X devices. However, this will be achieved through a dual-radio system --- one radio for C-V2X and another for NR V2X. The backward compatibility requirement for 802.11bd has implications on its design and performance, as we will discuss later in this paper.

IEEE 802.11bd and NR V2X are technologies that are currently under development, and hence, in this paper, we limit our discussions to a set of key features and functionalities that are likely to be included in the final standard. The main contributions of this paper are as follows: 
\begin{itemize}
    \item We outline the key objectives of 802.11bd and NR V2X, followed by a detailed description of important enhancements being made to DSRC and C-V2X in the process of development of 802.11bd and NR V2X, respectively.
    \item We elaborate on critical challenges encountered in the design of the two RATs and their potential solutions.
    \item We look at performance projections of 802.11bd and NR V2X in light of their respective design objectives.
    \item Finally, we discuss a number of key spectrum management issues that represent hurdles to the deployment and management of V2X technologies. 
\end{itemize}
While there are several interesting and challenging research problems in the design of these evolving RATs (e.g., V2X security), we restrict our focus in this paper to the design of the PHY and MAC layers. To the best of our knowledge, ours is the first work that provides an in-depth look at the design considerations and development process of these two evolutionary RATs. Throughout this paper, we use the terms DSRC and 802.11p interchangeably. A summary of acronyms used in this paper is outlined in Table~\ref{table:acroyms}.

\begin{table}[htb]
    \centering
    \caption{Summary of Acronyms}
    \label{table:acroyms}
    \begin{tabular}{|c|c|}
        \hline
        \textbf{Acronym} & \textbf{Full-form} \\ \hline
         3GPP & 3rd Generation Partnership Project \\
         BCC & Binary Convolutional Coding \\
         C-V2X & Cellular Vehicle-to-Everything \\ 
         DMRS & Demodulation Reference Signals \\ 
         DSRC & Dedicated Short Range Communications \\ 
         FCC & Federal Communications Commission \\
         FDM & Frequency Division Multiplexing \\
         ITS & Intelligent Transporartion Systems \\
         LDPC & Low Density Parity Check \\
         LTE & Long Term Evolution \\
         MAC & Medium Access Contol (layer) \\
         MCS & Modulation and Coding Scheme \\
         NLOS & Non line-of-sight \\
         OFDM & Orthogonal Frequency Division Multiplexing \\
         PDR & Packet Delivery Ratio \\
         PHY & Physical (layer) \\
         PSCCH & Physical Sidelink Control Channel \\
         PSFCH & Physical Sidelink Feedback Channel \\
         PSSCH & Physical Sidelink Shared Channel \\
         QAM & Quadrature Amplitude Modulation \\
         QoS & Quality of Service \\
         QPSK & Quadrature Phase Shift Keying \\ 
         RAT & Radio Access Technology \\
         SC-FDMA & Single Carrier Frequency Division Multiple Access \\
         TDM & Time Division Multiplexing \\
         UE & User Equipment \\ 
         U-NII & Unlicensed National Information Infrastructure \\
         V2X & Vehicle-to-Everything \\
         \hline
    \end{tabular}
    \vspace{-3mm}
\end{table}

\section{State-of-the-Art}

\subsection{Dedicated Short Range Communications (DSRC)}
The PHY and MAC layers of DSRC are defined in the IEEE 802.11p standard, which is largely derived from IEEE 802.11a. Traditionally, Wi-Fi standards have been developed for low mobility applications. However, since DSRC was designed for vehicular networks characterized by high-mobility, enhancements were introduced to make DSRC suitable for such environments. DSRC uses an Orthogonal Frequency Division Multiplexing (OFDM)-based PHY with a channel bandwidth of $10$~MHz. Thus, in comparison to Wi-Fi, DSRC sub-carrier spacing is reduced by a factor of two. 
The MAC protocol used in DSRC is Carrier Sense Multiple Access~\cite{hassan2011performance}. However, there is no exponential back-off in DSRC, i.e. the parameter Contention Window used in contention-based MAC protocols remains fixed in DSRC ~\cite{hassan2011performance} due to two main reasons, i) because DSRC is designed mainly for broadcast-based systems, there is no acknowledgement frame sent back to the transmitter\footnote{In traditional Wi-Fi networks, the contention window size is doubled when the transmitter does not receive an acknowledgement.}, and ii) exponential back-off can lead to large Contention Window sizes, thereby leading to high latencies. 


\subsection{Cellular V2X (C-V2X)}
\label{sec:v2x}
C-V2X is a V2X RAT developed by 3GPP in its Rel. 14. C-V2X users can benefit from leveraging the existing widespread cellular infrastructure. However, since the presence of cellular infrastructure cannot always be relied upon, C-V2X defines transmission modes that enable direct V2X communications using the \emph{sidelink} channel over the PC5 interface. 
3GPP Rel. 14 introduced two new sidelink transmission modes (modes 3 and 4) to support low latency V2X communications~\cite{phyprocedures}. 

The basic time-frequency resource structure of C-V2X is similar to that of LTE, i.e. the smallest unit of allocation in time is one sub-frame ($1$~msec comprising of $14$ OFDM symbols) and the smallest frequency-granularity is 12 sub-carriers of $15$~kHz each (i.e. $180$~kHz). In each OFDM sub-carrier, C-V2X devices can transmit using Quadrature Phase Shift Keying (QPSK) or 16-Quadrature Amplitude Modulation (QAM) schemes with turbo coding~\cite{molina2017lte}. In addition to data symbols, however, C-V2X users also transmit control information and reference signals. The demodulation reference signal (DMRS) is one such signal, which is used for channel estimation. In LTE, DMRS symbols are inserted in two of the fourteen OFDM symbols. However, since C-V2X is designed for high-mobility environments, four DMRS symbols are inserted in a C-V2X sub-frame~\cite{filippiieee802}. 

Because C-V2X can operate in in-coverage as well as out-of-coverage scenarios, C-V2X can operate using the traditional LTE air interface as well as the sidelink air interface. 

\subsubsection{V2X using LTE-Uu Air Interface}
\label{sec:v2x-uu-interface}
\emph{LTE-Uu} is the traditional air interface between an eNodeB and a User Equipment (UE). Any UE using the LTE-Uu interface must transmit its message to the eNodeB in the uplink, which is sent by the eNodeB to the destination UE in the downlink.
To reduce the scheduling overhead associated with V2X uplink transmissions, the eNodeB can use semi-persistent scheduling, whereby the eNodeB assigns resources to a UE not only for the very next transmission, but also for a number of subsequent transmissions. Semi-persistent scheduling is beneficial for V2X applications because a majority of such traffic is periodic and have similarly-sized packets~\cite{techreport}. 

\subsubsection{V2X using PC5 Air Interface}
\label{sec:v2x-pc5-interface}
The PC5 air interface enables direct communications between UEs without requiring every packet to pass through the eNodeB. UEs can use the PC5 interface both in the presence and absence of the eNodeB.

A transmitted packet on the PC5 interface comprises of the data component and the sidelink control information (SCI). The SCI carries important information required to decode the corresponding data transmission, such as the modulation and coding scheme (MCS) used, resources occupied by current and future transmissions, etc. The physical channel used to transmit the SCI is called the Physical Sidelink Control Channel (PSCCH), while the Physical Sidelink Shared Channel (PSSCH) carries the data component. In C-V2X, PSCCH and PSSCH are multiplexed in frequency, i.e. transmitted on different frequency resources in the same sub-frame. 

\textit{C-V2X sidelink Mode 3:}
In C-V2X sidelink mode 3, allocation of resources for sidelink transmissions is handled by the eNodeB. Naturally, this mode is defined for scenarios where eNodeB coverage is available.
The C-V2X sidelink mode 3 uses the following notable mechanisms.
\begin{itemize}
 \item Semi-persistent scheduling: Like in LTE-Uu, eNodeB supports semi-persistent scheduling for C-V2X mode 3. 
 \item UE-report based scheduling: UEs can report their observations on their radio environments to assist the eNodeB in sidelink resource allocation. 
 \item Cross-carrier scheduling: If an operator has two or more carriers at its disposal, the eNodeB can schedule resources on one of the carriers for sidelink transmissions over the other carrier(s). 
\end{itemize}

\textit{C-V2X sidelink Mode 4:}
UEs outside cellular coverage can use C-V2X sidelink mode 4, whereby UEs reserve resources autonomously using the \emph{resource reservation algorithm}. This resource reservation algorithm requires each UE to sense the channel for $1$~second and process the sensing results in order to ensure that neighboring UEs pick and reserve orthogonal (in time, frequency or both) resources semi-persistently, thereby minimizing packet collisions. We refer the interested reader to~\cite{molina2017lte} for more details on the resource reservation algorithm.

\section{Need for Evolution}
\label{sec:need}
\subsection{Performance of Existing Technologies}

\subsubsection{DSRC}
Over the years, DSRC has been extensively studied using analytical models~\cite{hafeez2013performance}, extensive simulation studies~\cite{yin2004performance} and field trials~\cite{chowdhury2018lessons}. 
It has been shown in~\cite{hassan2011performance} that DSRC performance is satisfactory for most vehicular safety applications that require the end-to-end latency to be around $100$~msec as long as the density of vehicles is moderate. If, however, the vehicular density exceeds a certain limit, DSRC performance rapidly deteriorates due to two major factors, (i) packet collisions due to simultaneous transmissions, and (ii) packet collisions due to hidden nodes.  
The poor scalability of DSRC is partially addressed by using congestion control mechanisms such as those standardized in~\cite{sae}. Such mechanisms typically involve the control of transmission parameters such as the transmission power or the message transmission rate (in number of packets/second) or both. 

\subsubsection{C-V2X} 
Compared to DSRC, C-V2X is a newer and less-studied technology. Most studies that characterize  C-V2X performance derive their results from simulation platforms. Reference~\cite{molina2017lte} shows that the performance of C-V2X sidelink mode 4 is superior to that of DSRC in terms of a higher link budget, which is corroborated through experimental findings in~\cite{fcc_adj}. Further, centralized control of resources in C-V2X sidelink mode 3 leads to an efficient utilization of the spectrum, thereby leading to better performance guarantees as demonstrated in~\cite{bazzi2017performance}. However, despite improvements over DSRC, when the traffic density increases, the performance of C-V2X, too, drops rapidly~\cite{molina2017lte}, particularly for C-V2X mode 4. The C-V2X mode 4 algorithm allows for frequency re-use over a given geographical area. When the traffic density increases, the re-use distance is reduced, resulting in an increased interference level among C-V2X users. 

\subsection{Nature of Supported Applications}
According to the results from past studies~\cite{hafeez2013performance, molina2017lte} and the QoS requirements set for safety applications~\cite{etsi-bsa}, DSRC and C-V2X are capable of supporting a basic set of vehicular safety applications that are based on issuing driver-alerts to indicate potentially dangerous situations. Most of these applications require the delivery of periodic messages and have requirements ranging from $1-10$~Hz peroidicity and $50-100$~msec end-to-end latency. Such applications are designed to aid the driver in driving safely and efficiently. These applications are referred to as day-1 applications due to the fact that V2X-capable vehicles are likely to support them before any of the advanced use-cases discussed in the next sub-section.

\subsection{Requirements of Advanced Vehicular Applications}
One obvious need for evolution of both RATs is to improve the reliability of existing use-cases while delivering packets under their latency budget. In addition to this, however, provisioning basic safety applications alone is unlikely to meet the requirements of self-driving autonomous cars. For example, while existing applications such as left turn assist and emergency electronic brake lights~\cite{cv2x-use-cases-no-req} are beneficial for vehicle safety, autonomous vehicles will require vehicles to be capable of transmitting messages indicative of maneuver changes, trajectory alignments, platoon formations, sensor data exchange, etc.~\cite{tr22886}. Besides, even for human-driven vehicles, processing of data received from sensors of surrounding vehicles---for example, where one vehicle shares its live camera feed with a vehicle behind it---is expected to increase the safety benefits well beyond what can be achieved by basic safety applications.

Requirements of some advanced vehicular applications have been studied by the 3GPP in~\cite{tr22886}. These advanced V2X use-cases not only improve road safety, but also assist in better traffic management and cater to the infotainment needs of passengers. These applications fall under four broad categories: (i) vehicle platooning, 
(ii) advanced driving, 
(iii) extended sensors, 
and (iv) remote driving. 
The QoS requirements of these applications are summarized in Table~\ref{table:requirements}.

\begin{table}[htb]
    \centering
    \caption{QoS requirements of advanced V2X applications}
    \label{table:requirements}
    \begin{tabular}{|p{1.3cm}|p{1cm}|p{1.0cm}|p{1.1cm}|p{1.0cm}|p{1.0cm}|}
    \hline
    \textbf{Use Case} & \textbf{Max.} & \textbf{Payload} & \textbf{Reliability} & \textbf{Data} & \textbf{Min.} \\ 
    \textbf{Group} & \textbf{Latency} & \textbf{Size} & \textbf{(\%)}& \textbf{Rate} & \textbf{Range} \\ 
     & \textbf{(msec)} & \textbf{(Bytes)} &  & \textbf{(Mbps)} & \textbf{(meters)} \\ \hline
    Vehicle Platooning & 10 - 500 & 50 - 6000 & 90 - 99.99 & 50 - 65 & 80 - 350 \\ \hline 
    Advanced Driving & 3 - 100 & 300 - 12000 & 90 - 99.999 & 10 - 50 & 360 - 500 \\ \hline
    Extended Sensors & 3 - 100 & 1600 & 90 - 99.999 & 10 - 1000 & 50 - 1000 \\ \hline
    Remote Driving & 5 & - & 99.999 & UL: 25 DL: 1 & - \\ \hline
    \end{tabular}
    \vspace{-3mm}
\end{table}

As shown in Table~\ref{table:requirements}, the latency and reliability requirements of these advanced V2X applications are much more stringent than those of basic safety applications. Furthermore, these advanced applications are characterized by the use of large and variable sized packets, and rely on messages that are transmitted aperiodically. This is in stark contrast to the applications that are based on the transmission of basic safety messages, which are transmitted periodically (typically once every $100$~msec). 
It is, therefore, clear that in order to support such diverse and challenging V2X applications, a major overhaul to the existing V2X technologies is necessary.

\section{IEEE 802.11bd: Evolution of IEEE 802.11p}
\label{sec:ngv}

\subsection{Objectives}
\label{sec:ngv-objectives}
During the development of IEEE 802.11p the focus was to develop a vehicular communication standard that assisted in (i) vehicular safety, (ii) better traffic management, and (iii) other applications that add value, such as parking and vehicular diagnostics. The requirements set for 802.11p were to support: 
\begin{itemize}
    \item relative velocities up to $200$~km/hr,
    \item response times of around $100$~msec, and
    \item communication range of up to $1000$~m.
\end{itemize}

The 802.11p standard derived its PHY and MAC layers from 802.11a. Since then, however, 802.11a has given way to its successors i.e., 802.11n and 802.11ac, while 802.11ax is in its final stages of standardization. Considering that 802.11p was developed nearly two decades ago, advanced PHY and MAC techniques introduced in 802.11n/ac/ax can be leveraged to enhance 802.11p. With this objective, the IEEE 802.11 Next Generation V2X Study Group was formed in March 2018. After an initial feasibility study, the IEEE 802.11bd Task Group was created in January 2019. The primary design objectives of 802.11bd include supporting the following~\cite{11-18-0861-09-0ngv}:

\begin{itemize}
    \item at least one mode that achieves twice the MAC throughput of 802.11p with relative velocities up to 500 km/hr;
    \item at least one mode that achieves twice the communication range of 802.11p; 
    \item at least one form of vehicle positioning in affiliation with V2X communications. 
\end{itemize}

Additionally, 802.11bd must support the following~\cite{11-18-0861-09-0ngv}:

\begin{itemize}
    \item Interoperability: 802.11p devices must be able to decode (at least one mode of) transmissions from 802.11bd devices, and vice-versa.
    \item Coexistence: 802.11bd must be able to detect 802.11p transmissions and defer channel access, and vice-versa.
    \item Backward compatibility: At least one mode of 802.11bd must be interoperable with 802.11p.
    \item Fairness: In co-channel scenarios, 802.11bd and 802.11p must get equal channel access opportunities.
\end{itemize}

\subsection{Mechanisms}
\label{sec:ngv-mechanisms}

\subsubsection{Midambles}
\label{sec:ngv-mechanisms-midambles}
The 802.11 PHY layer is OFDM-based with $64$ sub-carriers, typically with a sub-carrier spacing of $312.5$~kHz. The PHY layer of 802.11p was derived directly from that of 802.11a by reducing the sub-carrier spacing by a factor of two. For typical vehicle speeds, the $156.25$~kHz sub-carrier spacing provided a trade-off between multi-path fading and relative Doppler spread~\cite{cheng2008measurement}. Thus, one approach to designing the PHY of 802.11bd is to use the 802.11ac PHY as a base-line and half the sub-carrier spacing (denoted as ``$2 \times \textrm{down-clock}$'') so that the $64$ 802.11bd sub-carriers can fit in a $10$~MHz channel. However, it has been shown in~\cite{11-18-0513-02-0wng} that the 802.11ac PHY using $2 \times \textrm{down-clock}$ with Low Density Parity Check (LDPC) coding, in fact, performs inferior to that of 802.11p. This sub-par performance of the 802.11ac PHY is attributed to channel variations within the frame duration, resulting in the receiver's inability to decode the frame. 

To address the above issue, 802.11bd proposes to use \emph{midambles}, which are similar in form and function to the preamble except their location within the frame. 
The preamble, which is at the beginning of the frame, is used for initial channel estimation. However, for fast-varying channels, the initial estimate may quickly become obsolete. Midambles, which will be introduced in between the OFDM data symbols with appropriate frequency, will serve in channel tracking so that an accurate channel estimate is obtained for all data symbols. Note that in C-V2X and NR V2X, a similar role is played by DMRS symbols (see Sec.~\ref{sec:v2x}). 

\subsubsection{Re-transmissions}
\label{sec:ngv-mechanisms-retransmissions}
\begin{figure} [htb]
    \vspace{-2mm}
    \centering
    \includegraphics[scale=0.6, trim={7cm 2.5cm 13.3cm 5cm},clip, angle=-90]{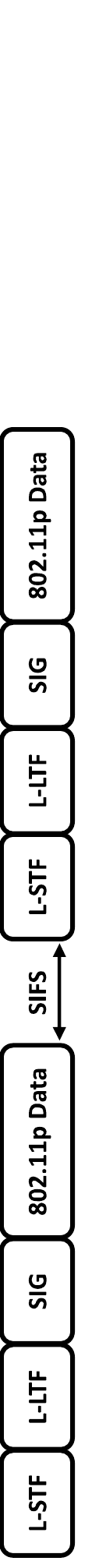}
    \caption{Frame format used for re-transmissions in 802.11bd.}
    \label{fig:frame-format-retransmission}
    \vspace{-2mm}
\end{figure}

A mechanism to increase the reliability is to have one or more re-transmissions of a packet. Using the frame structure shown in Fig.~\ref{fig:frame-format-retransmission}, reliability gains can be achieved for both 802.11p and 802.11bd devices. Note that for 802.11p devices, the original transmission and its re-transmission(s) appear as independent packets, and the packet is received successfully as long as one of the packet reception is successful. The initial transmission and its re-transmission(s) can either be sent within the same channel access opportunity or by using separate contention processes~\cite{11-18-1577-00-0ngv}. The TGbd proposes an adaptive re-transmission scheme, where decisions to re-transmit a frame and the number of re-transmissions are based on the congestion level~\cite{11-18-1186-00-0ngv}. A similar re-transmission mechanism is used in C-V2X to boost its reliability.

\subsubsection{Alternate OFDM Numerologies}
\label{sec:ngv-mechanisms-numerologies}
An OFDM symbol comprises of a cyclic prefix and the actual data symbol. The OFDM efficiency, i.e. the ratio of useful symbol duration to the total symbol duration, increases as the sub-carrier spacing decreases since the cyclic prefix duration is invariant of the symbol duration. To increase the OFDM efficiency, TGbd members are exploring the use of narrower OFDM numerologies (i.e. sub-carrier spacing) such that the number of sub-carriers is increased while still occupying a $10$~MHz channel~\cite{11-18-1553-00-0ngv}. These options include $2 \times \textrm{down-clock}$ with 64 sub-carriers, $4 \times \textrm{down-clock}$ with 128 sub-carriers, and $8 \times \textrm{down-clock}$ with 256 sub-carriers. The design of alternate OFDM numerologies must, however, take the maximum relative velocities into consideration. Channel variations across the frame duration can be estimated using midambles. However, if variations occur across an OFDM symbol, the resulting inter-carrier interference can be difficult to mitigate~\cite{11-18-1553-00-0ngv}.

\subsubsection{Dual Carrier Modulation}
\label{sec:ngv-mecanisms-dcm}
Dual Carrier Modulation (DCM) is a technique introduced in 802.11ax. DCM includes transmitting the same symbols twice over sufficiently far-apart sub-carriers such that frequency diversity is achieved~\cite{11-19-0009-00-00bd}. Because each symbol transmission is repeated over two different sub-carriers, the modulation order must be doubled (e.g. from BPSK to QPSK, or QPSK to 16-QAM) to maintain the throughput. Despite the increase in modulation order, DCM can help improve the block-error-rate (BLER) performance.

\subsubsection{Other PHY \& MAC Features}
\label{sec:ngv-mechanisms-other}
Other PHY layer features under consideration for inclusion in 802.11bd include the use of LDPC codes and multiple transmit/receive antennas to increase the reliability using spatial diversity or increase the throughput using spatial multiplexing~\cite{11-18-0860-03-0ngv, 11-18-0513-02-0wng}. 

At the MAC layer, to ensure equal and fair channel access opportunities for 802.11bd and 802.11p devices, 802.11bd will re-use 802.11p's contention parameters for different Enhanced Distributed Channel Access categories~\cite{11-19-0079-00-00bd}.

\subsubsection{mmWave Frequencies}
\label{sec:ngv-mechanisms-mmWave}
mmWave frequency bands (i.e., $60$~GHz and above) have enormous potential in catering to use-cases that require communication over small distances, but with a very high throughput (e.g., video streaming, downloading high-resolution 3D maps, etc). mmWave bands are particularly lucrative due to the abundance of spectrum therein, allowing very high throughputs even at lower order MCS~\cite{Microsof41:online}. 
The basis for design of mmWave 802.11bd can be existing 802.11 standards like 802.11ad, or its enhancement 802.11ay, which already operate in the mmWave bands~\cite{11-18-1187-03-0ngv}. However, one drawback of this frequency band is that its utility is limited to use-cases that do not require a large communication range.

\vspace{-4mm}



\subsection{Challenges}
\label{sec:ngv-challenges}
\subsubsection{Interoperability \& Backward Compatibility}
\label{sec:ngv-challenges-interoperability}
As described in Sec.~\ref{sec:ngv-objectives}, interoperability and backward compatibility are two critical requirements that 802.11bd must satisfy. Given that 802.11p equipped vehicles are already on the roads~\cite{filippiieee802}, without interoperability, 802.11bd (802.11p) devices will be able to communicate only with other 802.11bd (802.11p) devices, which is clearly undesirable. Interoperability and backward compatibility requirements place certain constraints on the design of PHY and MAC layers of 802.11bd. For example, multiple antenna schemes like space time block coding or the use of alternate waveforms violates the interoperability requirement~\cite{11-18-1186-00-0ngv}. 

\begin{figure}[H]
    \vspace{-2mm}
    \begin{center}
    \includegraphics[scale=0.55, trim={6.8cm 2.5cm 13.3cm 5cm},clip, angle=-90]{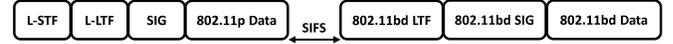}
    \caption{Interoperability through appending 802.11bd data}
    \label{fig:interoperability-append}
    \end{center}
    \vspace{-3mm}
\end{figure}

Fig.~\ref{fig:interoperability-append} shows a frame format that can be used by 802.11bd devices to achieve interoperability between 802.11bd and 802.11p devices~\cite{11-19-0082-01-00bd}. The three legacy fields in Fig.~\ref{fig:interoperability-append}, i.e. Legacy Short Training Field (L-STF), Legacy Long Training Field (L-LTF) and Legacy Signal (SIG), along with the 802.11p data field can be decoded by both 802.11p and 802.11bd. The same payload, but now using 802.11bd features such midambles, LDPC coding and higher order MCS (along with 802.11bd headers), is appended to the 802.11p data. During this extended duration, 802.11p devices will sense the channel as busy and defer channel access, while 802.11bd devices can achieve a higher reliability owing to the use of 802.11bd features and combining the two versions (i.e. 802.11p and 802.11bd) of the transmitted payload. 
The advantage of this packet structure is that its benefits are accrued without a need to change the higher layers -- a fundamental requirement for a smooth transition from 802.11p to 802.11bd.

Another frame format that achieves interoperability, while also increasing the reliability of 802.11bd transmissions is shown in Fig.~\ref{fig:interoperability-parity-concept}. The process of appending the parity bits to and extracting them from the transmitted data is shown in Fig.~\ref{fig:interoperability-parity-block}. The parity bits are generated by splitting the 802.11p data into blocks and encoding them using error control schemes such as the Reed Solomon (RS) code~\cite{11-18-1214-00-0ngv} (referred to as the outer code). The block comprising of 802.11p data and the RS parity bits are then encoded using the legacy Binary Convolutional Coding (BCC) scheme and appended at the end of the block. These parity bits can be leveraged at 802.11bd devices to increase the probability of successful decoding, while the contention state machine of legacy, i.e. 802.11p, devices ignore these parity bits~\cite{11-18-1214-00-0ngv}. To ensure that parity bits do not pass as valid data sequences, a single byte is added to the parity bits to make sure that the frame check sequence mechanism fails~\cite{11-18-1927-00-0ngv}.

\begin{figure}[htb]
         \centering
          \vspace{-3mm}
         \subfloat[Concept]{
         \includegraphics[scale=0.6, trim={9cm 2.7cm 11.1cm 15.8cm},clip, angle=-90]{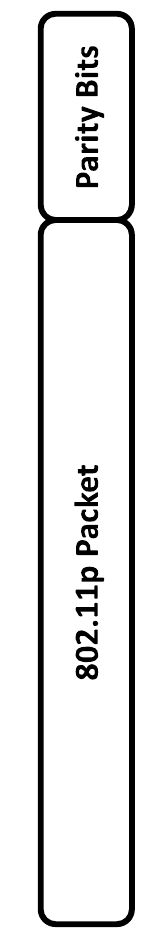}
         \label{fig:interoperability-parity-concept}
         } \\
         \subfloat[Implementation]{
         \includegraphics[scale=0.26, trim={0cm 4cm 0cm 3cm},clip]{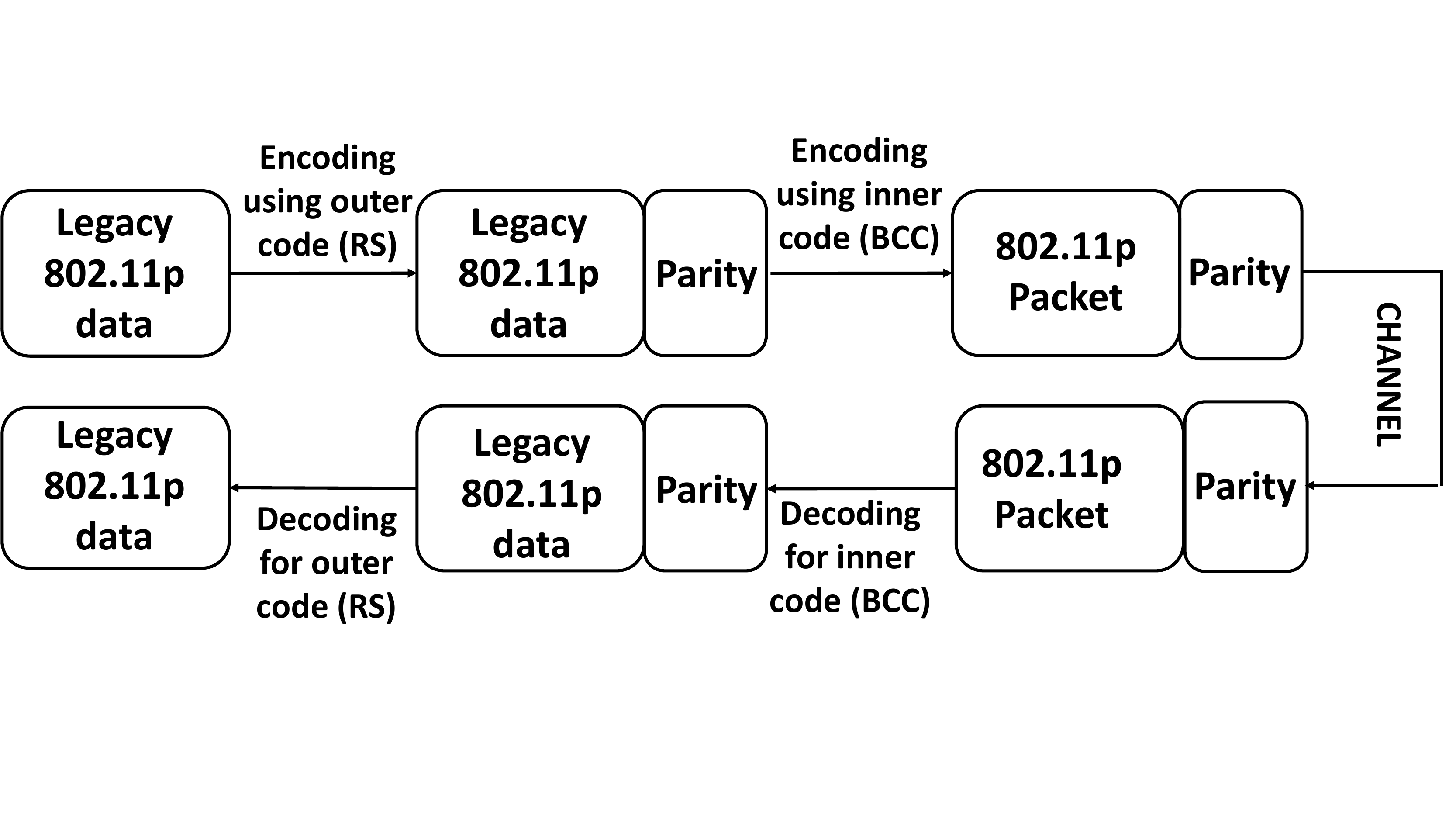}
         \label{fig:interoperability-parity-block}}
         \caption{Interoperability through appending parity bits}
  \label{fig:interoperability-parity}
  \vspace{-3mm}
\end{figure}

\subsubsection{Coexistence}
\label{sec:ngv-challenges-coexistence}
\begin{figure} [htb]
    \vspace{-2mm}
    \includegraphics[scale=0.6, trim={8.5cm 3.7cm 12cm 5cm},clip, angle=-90]{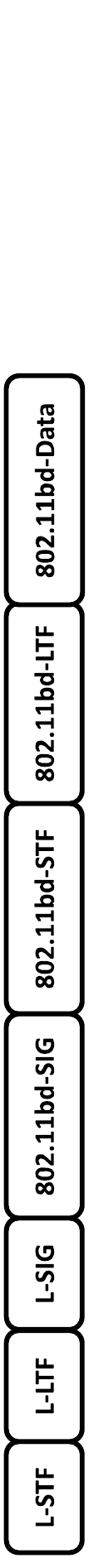}
    \caption{Frame format for 802.11p--802.11bd coexistence.}
    \label{fig:coexistence-802.11bd-802.11p}
    \vspace{-2mm}
\end{figure}

802.11bd also considers scenarios where coexistence between 802.11p and 802.11bd devices is desired. Coexistence differs from interoperability and backward compatibility in that the former does not require 802.11p devices to decode 802.11bd frames, but only to detect 802.11bd transmissions as valid 802.11 frames and defer channel access. Coexistence is desirable when the transmitted messages correspond to 802.11bd-specific use cases. Using the frame format shown in Fig.~\ref{fig:coexistence-802.11bd-802.11p}, 802.11bd devices transmit messages that are intended only for other 802.11bd (and not 802.11p) devices~\cite{11-19-0009-00-00bd}, while legacy devices will identify the channel as \emph{Busy} and defer channel access after decoding the legacy fields, i.e. L-STF, L-LTF and SIG. 

In a scenario where 802.11bd and 802.11p devices operate co-channel in the same geographical region, 802.11bd devices can certainly transmit all frames using the 802.11p frame format. However, there must be a mechanism for an 802.11bd device to notify other 802.11bd devices about its capabilities. Otherwise, it is possible that even after there are no 802.11p devices in the vicinity, 802.11bd devices continue to transmit using the 802.11p frame format. Several options exist for an 802.11bd to indicate its advanced capabilities. For example, the PHY or the MAC header can be used, with potential interoperability issues with the former option~\cite{11-19-0018-00-00bd-ngv}. 

\subsection{Performance Projections}
\label{sec:ngv-performance}
Link-level simulation results for 802.11bd have been reported in~\cite{11-18-1577-00-0ngv, 11-18-1214-00-0ngv, 11-19-0016-00-00bd, 11-18-0513-02-0wng}. Using the re-transmission scheme described in Sec.~\ref{sec:ngv-mechanisms-retransmissions} (see Fig.~\ref{fig:frame-format-retransmission}), 802.11bd devices can experience a gain of $3-8$~dB (at BLER of $10^{-1}$) by combining the original transmission and its re-transmission(s), while the corresponding gain for 802.11p devices is $0.5-1.7$~dB (at BLER of $10^{-1}$) depending on the number of re-transmissions used~\cite{11-18-1577-00-0ngv}. On the other hand, using the parity-based interoperability mechanism described in Sec.~\ref{sec:ngv-challenges-interoperability} (see Fig.~\ref{fig:interoperability-parity}), 802.11bd can benefit from a $1-3$~dB gain (at BLER of $10^{-1}$). Note, however, that for a given improvement in reliability of 802.11bd devices, the parity-based mechanism is more efficient because of the lower air-time utilization.

The DCM mechanism has been shown to provide gains of $4, 0.6$ and $2$~dB at MCS 0, MCS 1 and MCS 2, respectively in 802.11ax~\cite{11-15-1068-01-00ax}. DCM can be used in 802.11bd in coexistence scenarios where 802.11bd devices need to communicate only with other 802.11bd devices. In terms of throughput improvements, insertion of midambles in between data symbols (see Sec.~\ref{sec:ngv-mechanisms-midambles}) makes the use of higher order MCS feasible. This is illustrated in~\cite{11-19-0016-00-00bd}, where the throughput is shown to be doubled using midambles and LDPC coding over a $20$~MHz channel in the highway NLOS scenario. The doubling of throughput is observed only at large signal-to-noise ratio (SNR) values ($>20$~dB), which can be readily achieved in scenarios where the inter-vehicle distance is small---a typical scenario for high-throughput applications.  

The promising performance gains obtained at a single transmitter-receiver link, as discussed above, is expected to translate into a better system level performance. 
However, with the development of 802.11bd ongoing, the actual system-wide performance gain has not been demonstrated. 

\subsection{Comparison with IEEE 802.11p}
\label{sec:ngv-comparison}
Table~\ref{table:comparison-802.11p-802.11bd} summarizes the key differences between features/mechanisms of 802.11p and 802.11bd.

\begin{table}[htb]
 \centering
 \caption{Comparison of 802.11p and 802.11bd}
 \label{table:comparison-802.11p-802.11bd}
 \begin{tabular}{|p{2.9cm}|p{1.8cm}|p{2.7cm}|}
    \hline
    \textbf{Feature}  & \textbf{802.11p} & \textbf{802.11bd} \\ \hline
    Radio bands of operation & 5.9 GHz & 5.9 GHz \& 60 GHz \\ \hline
    Channel coding  & BCC & LDPC \\ \hline
    Re-transmissions & None & Congestion dependent \\ \hline
    Countermeasures against  & None  & Midambles \\
    Doppler shift & & \\ \hline
    Sub-carrier spacing & 156.25 kHz & 312.5 kHz, 156.25 kHz, 78.125 kHz \\ \hline
    Supported relative speeds & 252 kmph & 500 kmph  \\ \hline
    Spatial Streams & 1 & Multiple  \\ \hline
 \end{tabular}
 \vspace{-6mm}
\end{table}

\section{New Radio (NR) V2X: Evolution of C-V2X}
\label{sec:nrv2x}
\subsection{Objectives}
\label{sec:nrv2x-objectives}
The NR V2X Study Item~\cite{RP-181480} indicates that the design objective of NR V2X is \emph{not} to replace C-V2X, but to supplement C-V2X in supporting those use cases that cannot be supported by C-V2X. Because C-V2X is already standardized and commercial deployments are underway~\cite{ford}, it is likely that C-V2X and NR V2X might coexist in the same geographical region, where newer vehicles will have both C-V2X and NR V2X capabilities. Under such circumstances, use cases that can be supported reliably by using C-V2X can use C-V2X procedures, while the remaining use cases can use NR V2X procedures~\cite{RP-181480}. However, to ensure that NR V2X can provide a unified support for all V2X applications in the future, NR V2X must be capable of supporting not only advanced V2X applications, but also basic safety applications that are supported by present-day C-V2X. 

NR V2X is being designed to support V2X applications that have varying degrees of latency, reliability and throughput requirements. While some of these use cases require the transmission of periodic traffic, a large number of NR V2X use cases are based on reliable delivery of aperiodic messages. Furthermore, while some use cases require broadcast transmissions, others such as vehicle platooning are efficiently supported by transmission of messages only to a specific sub-set of vehicles (UEs). In some cases, in fact, 3GPP sees benefits in transmitting packets to only a single vehicle (UE)~\cite{tr22886}. To support such use cases, two new communication types, viz., unicast and groupcast, will be introduced in NR V2X. Like IEEE 802.11bd, NR V2X also considers the use of mmWave bands for V2X applications, particularly for applications that require a short range and high to very high throughputs. However, considering the limited timeline of Rel. 16 (expected to be standardized by Dec. 2019), NR V2X mmWave operations are de-prioritized in its Study Item.

The NR V2X Study Item outlines its following objectives. 
\begin{itemize}
    \item \emph{Enhanced sidelink design:} Re-design sidelink procedures in order to support advanced V2X applications.
    \item \emph{Uu interface enhancements:} Identify enhancements to the NR Uu interface to support advanced V2X applications.
    \item \emph{Uu interface based sidelink allocation/configuration:} Identify enhancements for configuration/allocation of sidelink resources using the NR Uu interface. 
    \item \emph{RAT/Interface selection:} Study mechanisms to identify the best interface (among LTE sidelink, NR sidelink, LTE Uu and NR Uu) for a given V2X message transmission.
    \item \emph{QoS Management:} Study solutions that meet the QoS requirements of different radio interfaces.
    \item \emph{Coexistence:} Feasibility study and technical solutions for coexistence of C-V2X and NR V2X within a single device, also referred to as \emph{in-device coexistence}.
\end{itemize}

\subsection{Terminology}
\label{sec:nrv2x-terminology}
\subsubsection{NR V2X sidelink modes}
\label{sec:nrv2x-terminology-modes}
Like C-V2X, NR V2X defines two sidelink modes. The \emph{NR V2X sidelink mode 1} defines mechanisms that allow direct vehicular communications within gNodeB coverage. In this mode, the gNodeB allocates resources to the UEs. The \emph{NR V2X sidelink mode 2}, on the other hand, supports direct vehicular communications in the out-of-coverage scenario. For direct comparison with 802.11bd, in the remainder of this section, we keep our discussions restricted to NR V2X sidelink mode 2.

\subsubsection{Unicast, Groupcast and Broadcast}
\label{sec:nrv2x-terminology-comm-types}
\begin{figure}
    \centering
    \includegraphics[scale=0.35, trim={6cm 4.7cm 6cm 0cm}, clip, angle=-90]{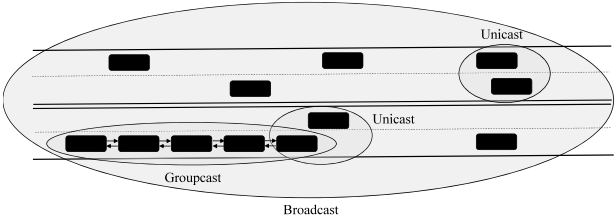}
    \caption{Communication types in NR V2X.}
    \label{fig:comm-types}
    \vspace{-5mm}
\end{figure}

In NR V2X unicast transmissions, the transmitting UE has a \emph{single} receiver UE associated with it. On the other hand, the groupcast mode is used when the transmitting UE wishes to communicate with more than one, but only a specific sub-set of UEs in its vicinity. Finally, broadcast transmissions enable a UE to communicate with all UEs within its transmission range. These communication types are illustrated in Fig.~\ref{fig:comm-types}. Note that C-V2X only provisions support for broadcast transmissions. A single UE can have multiple communication types active simultaneously. For example, a platoon leader UE can communicate with its platoon member UEs using the groupcast mode, while using the broadcast mode to transmit other periodic messages to UEs that are not part of the platoon as shown in Fig.~\ref{fig:comm-types}. 

\vspace{-4mm}

\subsection{Mechanisms}
\label{sec:nrv2x-mechanisms}

\subsubsection{Use of NR Numerologies}
\label{sec:nrv2x-mechanisms-numerologies}
Support for flexible numerologies is a key feature introduced in 3GPP Rel. 15. As opposed to a fixed sub-carrier spacing used in LTE, NR supports different sub-carrier spacings, which are multiples of the LTE sub-carrier spacing, i.e. 15 kHz. Sub-carrier spacing of 15, 30 and 60 kHz will be supported for sub-6 GHz NR V2X (i.e. Frequency Range 1, FR1), while 60 and 120 kHz will be supported for frequency bands above 6 GHz (i.e. FR2)~\cite{R1-1810051}. 
The use of higher sub-carrier spacings facilitates latency reduction. Assuming each UE requires one slot for its transmission, the transmission time of a UE decreases as the sub-carrier spacing increases. Note that the term ``slot'' and ``sub-frame'' hold different meanings in NR V2X. NR defines the duration corresponding to 14 OFDM symbols as one slot, while a sub-frame has a fixed duration of 1 msec. 
Furthermore, due to smaller slot durations at higher sub-carrier spacings, variations within the slot will be smaller, thereby needing fewer DMRS symbols per slot for channel estimation.

\subsubsection{Slot, Mini-slot and Multi-slot Scheduling}
\label{sec:nrv2x-mechanisms-scheduling}
In LTE and C-V2X, the transmission time is tightly coupled to the sub-frame duration, i.e. all UEs always transmit for a duration of 1 sub-frame (1 msec). However, if a UE has only a small amount of data to send, which can be accommodated in less than 14 OFDM symbols, allocating the entire slot for its transmission is resource inefficient. Second, whenever a packet arrives at a UE for transmission, the UE has to wait until the beginning of the next slot to begin transmitting. Such a slot-based scheduling is supported by default in NR V2X. However, NR V2X will also support \emph{mini-slot} scheduling, where UEs that have latency-critical messages to send can start their transmissions at any of the 14 OFDM symbols and can occupy any number of OFDM symbols within the slot. 
Furthermore, slot-aggregation, i.e. combining two or more slots to form a \emph{multi-slot}, will also be supported in NR V2X to cater to use-cases that require exchange of large-sized packets. 


\subsubsection{Multiplexing of PSCCH and PSSCH}
\label{sec:nrv2x-mechanisms-multiplexing}
\begin{figure}
         \centering
                 \subfloat[Multiplexing in C-V2X]{
                 \includegraphics[scale=0.35, trim={6.5cm 2.5cm 6cm 16cm},clip,angle=-90]{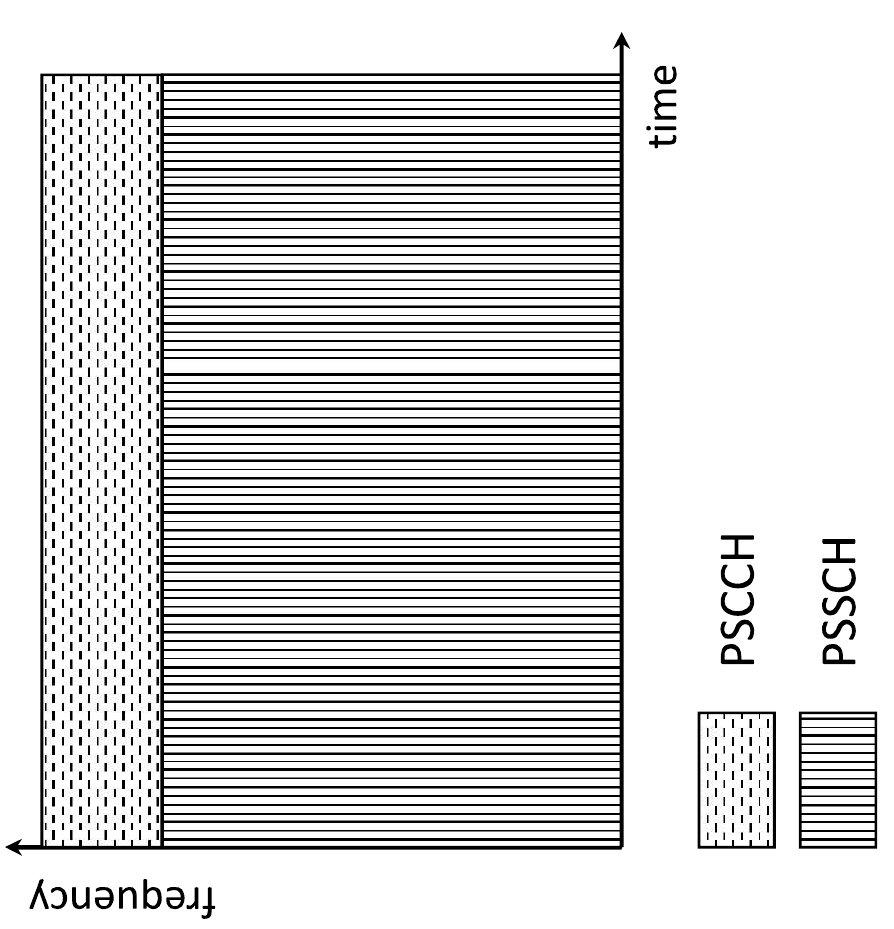}
                 \label{fig:multiplexing-cv2x}}
                 \subfloat[Multiplexing in NR V2X]{
                 \includegraphics[scale=0.35, trim={6.5cm 2.5cm 6cm 16cm},clip,angle=-90]{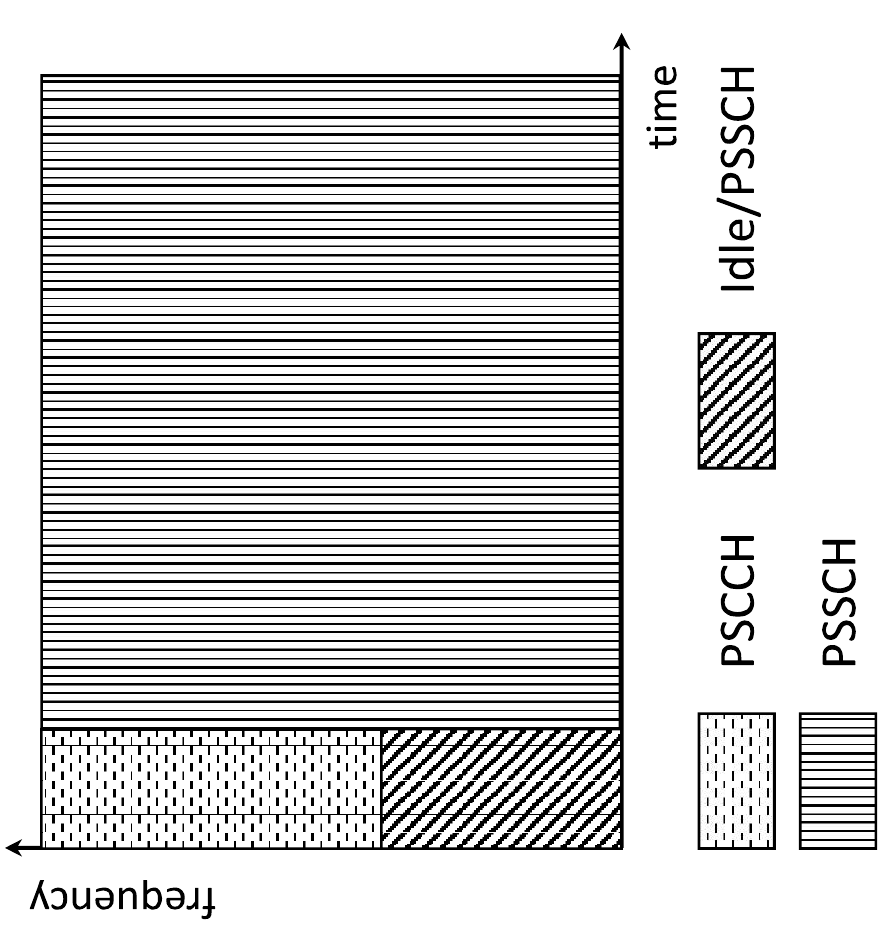}
                 \label{fig:multiplexing-nrv2x}}
         \caption{Multiplexing of PSCCH and PSSCH in C-V2X and NR V2X.}
  \label{fig:multiplexing}
  \vspace{-4mm}
\end{figure}

In C-V2X, PSCCH and PSSCH are multiplexed in frequency (see Fig.~\ref{fig:multiplexing-cv2x}). The drawback of this approach is that a receiver must buffer the message for the entire sub-frame and can decode the message only at the end of the sub-frame. This may prove to be inefficient in NR V2X due to tight latency constraints of certain messages. To address this problem, PSCCH and PSSCH will be multiplexed in time in NR V2X, i.e. the PSCCH will be transmitted first, followed by the transmission of PSSCH. This is illustrated in Fig.~\ref{fig:multiplexing-nrv2x}, where the use of resources marked as ``Idle/PSSCH'' is still under consideration and can be left idle or used for the transmission of PSSCH~\cite{R1-1810051}. 



\subsubsection{Introduction of sidelink feedback channel}
\label{sec:nrv2x-mechanisms-feedback}
For unicast and groupcast communications, reliability can be improved if the source UE can re-transmit the packet if the reception fails at UE(s) during the initial transmission. Although C-V2X provides support for re-transmissions, these re-transmissions are blind, i.e. the source UE, if configured, re-transmits without knowing if the initial transmission has been received by surrounding UEs. Such blind re-transmissions are resource inefficient if the initial transmission is successful. Blind re-transmissions are also ineffective if more than two transmissions are required for a given reliability requirement. Furthermore, if the source UE has access to the channel state information at its destination UE, this can be leveraged to adapt transmission parameters such as the MCS. To facilitate these two enhancements, i.e. enabling feedback-based re-transmissions and channel state information acquisition, NR V2X will introduce a new feedback channel---Physical Sidelink Feedback Channel (PSFCH)~\cite{R1-19xxxxx}. Selection of resources to use for PSFCH is still under study. However, in order to reduce the complexity associated with resource selection, preliminary studies in 3GPP recommend that the transmitter UE must notify the receiver UE about which resource to use for transmission on PSFCH~\cite{R1-1903366}. 

\subsubsection{Other PHY layer enhancements}
\label{sec:nrv2x-mechanisms-other}
In addition to the above features, NR V2X will include many other enhancements at the PHY layer, most of which are inherited from NR. These include the use of LDPC coding, higher order MCS including 64-QAM, and a flexible number of DMRS symbols per slot. 

\subsubsection{Introduction of new sub-modes of NR sidelink mode 2}
\label{sec:nrv2x-mechanisms-submodes}
Unlike in C-V2X sidelink mode 4, where there were no sub-modes, 3GPP began with evaluating four sub-modes of NR V2X sidelink mode 2~\cite{R1‑1809867}. These sub-modes are as follows:

\begin{itemize}
    \item Mode 2 (a): Each UE autonomously selects its resources. This mode is similar to C-V2X sidelink mode 4. 
    \item Mode 2 (b): UEs assist other UEs in performing resource selection. The UE providing assistance can be the receiver UE, which can potentially notify the transmitting UE of its preferred resources using the PSFCH.
    \item Mode 2 (c): In this sub-mode, UEs use pre-configured sidelink grants to transmit their messages. This sub-mode will be facilitated through the design of two-dimensional time-frequency patterns such as those described in~\cite{R1-1812209}.
    \item Mode 2 (d): UEs select resources for other UEs. 
\end{itemize}

In subsequent 3GPP meetings, it has been agreed to no longer support modes 2(b) and 2(c) as separate sub-modes~\cite{R1-1903397}. UE assistance (i.e., mode 2(b)) can be used in modes 2(a)/(d) improve the performance of resource selection. On the other hand, even though mode 2 (c) will not be supported as a separate sub-mode, the use of time-frequency resource patterns in mode 2(a) is not precluded~\cite{R1-1903397}. 
The design of modes 2(a) and 2(d) present unprecedented challenges, which are discussed through the rest of this sub-section.

\textit{Design of mode 2 (a):}
In this sub-mode, the transmitting UE must select its resources in an autonomous fashion. The UE can use any assistance information facilitated by sub-mode 2 (b). However, this is unlikely at least for broadcast transmissions because gathering information from all receiving UEs will lead to a prohibitively high overhead. 

The C-V2X sidelink mode 4 resource reservation algorithm leverages the periodicity and fixed-size assumption of basic safety messages. Since this assumption is no longer valid for NR V2X use-cases in general, the resource selection mechanism must be re-engineered. For aperiodic traffic, since the arrival of future packets cannot be inferred from sensing previous transmissions from surrounding UEs, several 3GPP members propose the use of \emph{short-term sensing}~\cite{R1-1812491}. A classic example of short-term sensing is the Listen Before Talk protocol, such as what is used in Wi-Fi or in License Assisted Access\footnote{License Assisted Access is the unlicensed flavor of LTE developed by 3GPP in its Rel. 14 for operations in the unlicensed frequency bands.}. 
On the other hand, periodic traffic can be supported using \emph{long-term sensing} similar to that used in C-V2X. Long term sensing involves sensing and analyzing the channel occupation during the sensing window, and using this information to select resources from the selection window. Some suggested enhancements to the C-V2X sidelink mode 4 algorithm are as follows~\cite{R1-1812491}: 

\begin{itemize}
 \item Make the duration of the sensing window flexible based on vehicular mobility in contrast to the fixed $1$~sec window used in C-V2X mode 4. This is beneficial in high-mobility scenarios, where sensing results can quickly become obsolete.
 \item It has been shown in~\cite{R1-1812491} that by eliminating the RSSI averaging procedure used in the C-V2X mode 4 resource reservation algorithm (see~\cite{molina2017lte} for details), the performance of long-term sensing can be improved.
\end{itemize}

\begin{figure}[htb]
\vspace{-2mm}
\centering
\includegraphics[scale=0.45, trim={3.5cm 4cm 4cm 8cm},clip,angle=-90]{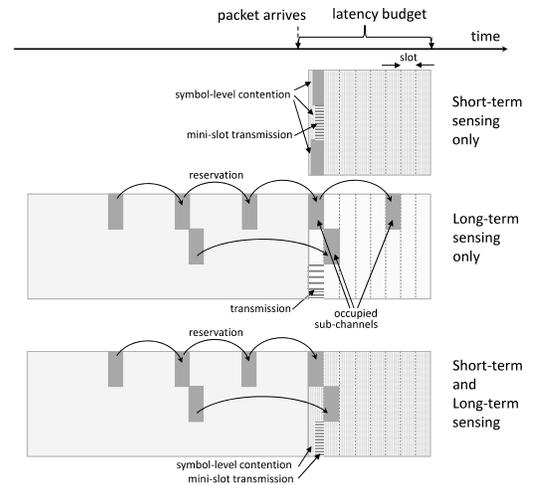}
\caption{Short-term and long-term sensing in NR V2X. 
}
\label{fig:sensing}
\end{figure}

The short-term and long-term sensing procedures can be used as stand-alone mechanisms in sub-mode 2 (a), or the two can be used as supplementary mechanisms. Reference~\cite{R1‑1902484} describes how long-term and short-term sensing can be used in conjunction. If both long-term and short-term sensing are configured, then UEs perform sensing and resource exclusion over the sensing window and select a transmission resource within the selection window. However, before transmitting, the UE must perform short-term sensing to detect the presence of other signals on its selected resource. This is beneficial for scenarios where periodic and aperiodic traffic will share the same resource pool(s). The use of short-term, long-term and combined sensing is illustrated in Fig.~\ref{fig:sensing}.

\textit{Design of mode 2 (d):} 
In this sub-mode, a UE performs resource allocation for a group of UEs in its vicinity. This sub-mode is especially useful in platooning applications where vehicles move along the same direction with small relative velocities~\cite{R1-1810977}. Furthermore, this sub-mode is likely to be used in applications that require groupcast or unicast transmissions. The UE performing resource allocation for other UEs within the group is referred to as the \emph{scheduling UE} (S-UE)~\cite{R1-1814260}. Because the S-UE can overlook the resource allocation of UEs within its group, this mode is beneficial in significantly reducing the number of collisions between group member UEs. 

The mechanism to select a UE as the S-UE is still under study. Some possible options include selection based on geo-location or pre-configuration. Geo-location based selection of S-UE is beneficial in platooning applications, where the length of platoon can be as long as $250$~m~\cite{R1-1901844}. In such conditions, a vehicle at the center of the platoon is more likely to have an accurate estimate of radio environments of all vehicles in the platoon than vehicles at the front or back of the platoon. Pre-configuration based S-UE selection, on the other hand, implies that certain vehicles with additional hardware/processing capabilities can take on the responsibility of resource allocation for surrounding vehicles. Regardless of the eventual mechanism used to select S-UE, the selection procedure will be performed at the application layer~\cite{R1-1903397}. 

Despite its benefits, this sub-mode's success will depend on the efficiency of solutions to the following sub-problems: 
\begin{itemize}
    \item Which resources can the S-UE use while performing resource allocation for its member UEs? In particular, can the S-UE coexist in a resource pool where other UEs select resources using mode 2 (a)?
    \item How to mitigate interference between neighboring UEs that are assigned resources by different S-UEs?
    \item Once the S-UE performs resource allocation for its member UEs, how does it convey these allocations to the member UEs, without prohibitively increasing the signaling overhead?
\end{itemize}

\subsection{Challenges}
\label{sec:nrv2x-challenges}
\subsubsection{Coexistence of C-V2X and NR V2X}
\label{sec:nrv2x-challenges-coexistence-rats}
Vehicles equipped with C-V2X are expected to hit the roads soon~\cite{ford}. Considering that vehicles typically have a life-span of one or more decades~\cite{RP-180690}, NR V2X is likely to have to coexist with C-V2X. However, NR V2X is not backward compatible with C-V2X~\cite{RP-181480}. 
This incompatibility stems from, among other factors, the use of multiple numerologies in NR V2X. A C-V2X device operating at 15 kHz sub-carrier spacing, cannot decode messages transmitted using the 30 or 60 kHz spacing. Newer vehicles will, thus, be equipped with modules of both technologies, i.e. C-V2X and NR V2X, making it imperative to design effective coexistence mechanisms~\cite{RP-181480}.

For C-V2X and NR V2X coexistence, the NR V2X study item~\cite{RP-181480} considers only the ``not co-channel'' scenario, i.e. a scenario where C-V2X and NR V2X operate in different channels. Two approaches can be used for such non co-channel coexistence~\cite{R1-1812000}: i) frequency division multiplexing (FDM), or ii) time division multiplexing (TDM). Note that the term TDM is somewhat misleading in this context because not only are C-V2X and NR V2X resources orthogonal in time, but they are also orthogonal in frequency.

\textit{FDM approach for coexistence:} In this approach, transmissions on the two RAT can overlap in time. This approach is advantageous because there is no need for tight time synchronization between the C-V2X and NR V2X modules. However, despite the use of two different radios (one for C-V2X and another for NR V2X), if the assigned channels are adjacent or are not sufficiently far apart, leakage due to out-of-band emissions from one radio terminal will impair the reception at the other radio terminal. Furthermore, if the two RATs operate in the same band (e.g. 5.9 GHz ITS band), the total power radiated by the vehicle may be restricted by regulatory limits and may have to be split across the two RATs, affecting the QoS requirements of V2X applications. 

\textit{TDM approach for coexistence:} In this approach, transmissions on the two RATs occur in different channels \emph{and} at different time instants. The advantage of this approach is that the maximum permissible transmission power can be used by both technologies because only one interface transmits at any given time. Furthermore, there is no leakage across channels. However, the TDM approach is disadvantageous for latency critical use-cases as the NR V2X interface may be \emph{off} while a latency sensitive packet is generated at the vehicle. Additionally, the TDM approach puts severe restrictions on time synchronization between C-V2X and NR V2X~\cite{R1-1812624}. 

\subsubsection{Coexistence across communication types \& periodicities}
\label{sec:nrv2x-challenges-coexistence}

Different messages transmitted by the same UE using NR V2X may have very different QoS requirements. For example, the platoon leader UE in Fig.~\ref{fig:comm-types} must transmit broadcast, groupcast and unicast messages. Furthermore, some of these messages may be periodic, while others aperiodic. Periodic broadcast traffic can use the C-V2X sidelink mode 4 resource reservation algorithm in the out-of-coverage scenario. However, other categories of traffic (such as aperiodic unicast) may use different transmission mechanisms as discussed in Sec.~\ref{sec:nrv2x-mechanisms}. 

\begin{figure}[htb]
\vspace{-2mm}
\centering
\includegraphics[scale=0.6, trim={7.7cm 4.2cm 7cm 11cm},clip,angle=-90]{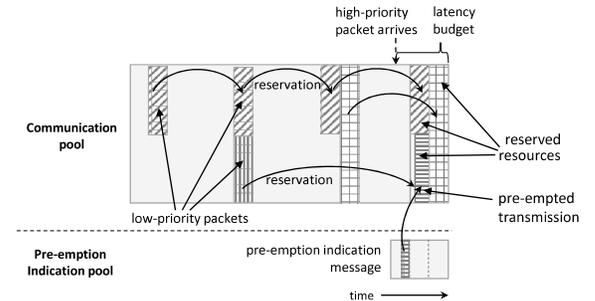}
\caption{Pre-emption mechanism in NR V2X}
\label{fig:pre-emption}
\vspace{-3mm}
\end{figure}

One approach to address the coexistence issue between diverse traffic patterns is to use the \emph{pre-emption} mechanism~\cite{R1-1810977}. Consider a scenario where periodic and aperiodic traffic share the same resource pool(s) while a high priority message arrives at a UE and all resources within the packet's latency budget are reserved by low-priority periodic traffic. Using the pre-emption mechanism, the UE can use one of the resources that was originally reserved by another UE for lower-priority traffic as illustrated in Fig.~\ref{fig:pre-emption}. The UE's intent to pre-empt such traffic can be communicated through the use of a pre-emption indication (PI) message---a message that can be sent on the same resource pool or on a dedicated resource pool reserved for transmitting PI messages. 

Design of the pre-emption mechanism is, however, not devoid of challenges. For example, frequent pre-emptions of periodic traffic may significantly affect the reliability of applications that rely on such messages. A resource inefficient, yet simple, alternative to pre-emption is to use separate resource pools, one for each class of communication requirements~\cite{R1-1810485}. 

\subsection{Performance Projections}
\label{sec:nrv2x-performance}
 Although the development of mechanisms that will constitute NR V2X is ongoing, preliminary performance studies have been conducted by some 3GPP members~\cite{R1-1812210, R1-1812211, R1-1813230}. Results in~\cite{R1-1813230} indicate that large performance gains can be achieved using the $60$~kHz sub-carrier spacing over the $15$~kHz spacing used in C-V2X. These gains are more pronounced at higher relative velocities (280, 500 kmph). For example, at 500 kmph, using 4 DMRS symbols/slot and QPSK modulation, a coding gain of $7$~dB is achieved for a BLER of $10^{-2}$. In order to cover large distances, use of the $60$~kHz sub-carrier spacing requires the extended cyclic prefix, adding to the communication overhead. However, this can be compensated for by using fewer DMRS symbols/slot at $60$~kHz sub-carrier spacing. As shown in~\cite{R1-1812211}, with the use of multiple-antenna techniques, reduction in the number of DMRS symbols/slot from 4 to 2 at $60$~kHz sub-carrier spacing leads to practically no loss even at 500 kmph! Thus, by leveraging larger sub-carrier spacings made possible by the use of flexible NR numerologies, NR V2X can significantly outperform C-V2X.
 
 The superior link-level performance of NR V2X translates to a superior system-level performance as shown in~\cite{R1-1812210}. Under highway scenarios, using the $60$~kHz sub-carrier spacing and a $20$~MHz channel, the packet delivery rate (PDR) is $\sim99.7-99.8\%$ for all communication types (i.e. broadcast, groupcast and broadcast) and message types (i.e. periodic and aperiodic). The packet generation rate for periodic traffic is $10$~Hz, while for aperiodic traffic a packet is generated once every $X$~msec, where $X = (50+\textrm{ an exponential random variable with mean 50})$~msec. This indicates that, at least in the highway scenario, NR V2X is close to meeting some of the performance requirements outlined in Table~\ref{table:requirements}. However, in the urban scenario, which is typically characterized by a higher density of vehicles and large path losses, the performance of NR V2X varies in $\sim93-97\%$ range, thereby indicating that further enhancements are required for reliable communications in urban environments. Further, results presented in~\cite{R1-1812210} are for relatively low message transmission rates ($\sim10$~Hz). The performance of NR V2X for more demanding applications remains to be seen. It must be noted that results shown in~\cite{R1-1812210, R1-1812211, R1-1813230} do not account for all features described in Sec.~\ref{sec:nrv2x-mechanisms} and with the development and refinement of several mechanisms in NR V2X still underway, the gap between performance requirements and achieved performance may become smaller by the time NR V2X is finalized.
 
 \vspace{-3mm}
 \subsection{Comparison with C-V2X}
 \label{sec:nrv2x-comparison}
 Table~\ref{table:comparison-cv2x-nrv2x} summarizes the key differences between features/mechanisms of C-V2X and NR V2X.
 
 \begin{table}[htb]
     \centering
     \caption{Comparison of C-V2X and NR V2X}
     \label{table:comparison-cv2x-nrv2x}
     \begin{tabular}{|p{2.25cm}|p{1.9cm}|p{3.55cm}|}
        \hline
        \textbf{Feature}  & \textbf{C-V2X} & \textbf{NR V2X} \\ \hline
        Comm. types & Broadcast & Broadcast, Groupcast, Unicast \\ \hline
        MCS & QPSK, 16-QAM & QPSK, 16-QAM, 64-QAM \\ \hline
        Waveform & SC-FDMA & OFDM \\ \hline
        Re-transmissions & Blind & HARQ \\ \hline
        PHY channels & PSCCH, PSSCH & PSCCH, PSSCH, PSFCH \\ \hline
        Control \& data multiplexing & FDM & TDM \\ \hline
        DMRS & Four/sub-frame & Flexible \\ \hline
        Sub-carrier spacing & 15 kHz & sub-6 GHz: 15, 30, 60 kHz \\ 
        & & mmWave: 60, 120 kHz \\ \hline
        Scheduling interval & one sub-frame & slot, mini-slot or multi-slot \\ \hline
        Sidelink modes & Modes 3 \& 4 & Modes 1 \& 2 \\ \hline
        Sidelink sub-modes & N/A & Modes 2(a), 2(d) \\ \hline
     \end{tabular}
     \vspace{-5mm}
 \end{table}
\section{Spectrum Management Issues}
\label{sec:spectrum-management}

\subsection{Interoperability of multiple V2X RATs}
\label{sec:spectrum-management-interoperability}
With DSRC and C-V2X as the two major developed RATs, and their corresponding evolutions---IEEE 802.11bd and NR V2X---underway, regional regulators and automakers have different choices to provision V2X communications in vehicles. While some automakers have committed to the use of DSRC in their vehicles~\cite{toyota}, others prefer C-V2X~\cite{ford}. At present, only DSRC is permitted to operate in the ITS band in US, while the 5G Automotive Association has requested a waiver to the US Federal Communications Commission (FCC) to allow C-V2X operations in the 5905-5925 MHz~\cite{5gaaWaiver}. The US Department of Transportation subsequently released a request for comments~\cite{dot_rfc} on questions regarding the choice of V2X RAT, their interoperability, and ability to support advanced V2X applications. Considering these factors, at least in the initial years, it is plausible that multiple V2X technologies may operate simultaneously within a given geographical region.


C-V2X and DSRC are not compatible with each other. Thus, if some vehicles use DSRC and others use C-V2X, these vehicles will be unable to communicate with each other---a scenario where the true potential of V2X communications cannot be attained. 
In order to address this, some proposals such as~\cite{kenney_bd} suggest that, at least within a given geographical region, regulatory agencies must permit only one V2X technology (either DSRC or C-V2X) to operate in a vehicle.

\vspace{-2mm}
\subsection{Coexistence with Wi-Fi}
\label{sec:spectrum-management-coexistence}
In the US and Europe, the 5.9 GHz band has been explored for Wi-Fi-like secondary operations. Because V2X applications demand high reliability, unlicensed Wi-Fi operations can be permitted only if they do not cause interference to the primary V2X technologies. At the time regional regulators started studying the possible coexistence of Wi-Fi and V2X technologies, DSRC was the only developed V2X technology. Therefore, solutions for secondary Wi-Fi operations in US and Europe~\cite{website:Channelization, tr103319} have been developed considering DSRC as the default V2X technology. Besides such standardized solutions, the study on coexistence between DSRC and Wi-Fi has received considerable attention in the literature~\cite{park2014coexistence, naik2017coexistence, liu2017coexistence}. A large number of such studies leverage the similarities in the MAC protocols of Wi-Fi and DSRC. Thus, by increasing the Contention Window size and/or Inter-frame space of Wi-Fi, the priority of Wi-Fi transmissions can be reduced so that if packets are available at both DSRC and Wi-Fi transmitters, DSRC transmitters will gain access to the channel with a higher probability. Since the MAC protocol of 802.11bd is expected to be similar to that of 802.11p~\cite{11-19-0079-00-00bd}, it is possible that coexistence mechanisms developed for DSRC--Wi-Fi coexistence may also be suitable for 802.11bd--Wi-Fi coexistence. 

In contrast to DSRC--Wi-Fi coexistence, the coexistence between C-V2X and Wi-Fi has not been well investigated. This subject has only been touched upon in~\cite{naik2018coexistence}. It must be noted that mechanisms developed for DSRC and Wi-Fi coexistence cannot be re-used for C-V2X and Wi-Fi coexistence because C-V2X uses a considerably different MAC protocol from Wi-Fi. If C-V2X is used to provision V2X applications, solutions for secondary Wi-Fi operations need to be re-designed with the MAC protocol of C-V2X taken into consideration. Ideally, coexistence mechanisms must be agnostic to the choice of V2X technology. However, considering the differences in MAC protocols of DSRC and C-V2X, a unified coexistence mechanism may be improbable. Nevertheless, mechanisms must be at least forward compatible, i.e. any coexistence mechanism developed for C-V2X--Wi-Fi coexistence must also be compatible with NR V2X--Wi-Fi coexistence. 

\vspace{-3mm}
\subsection{Interference from Adjacent Bands}
\label{sec:spectrum-management-interference}
The performance of the V2X RAT can be affected by out-of-band emissions from adjacent bands. In the US, at the lower end of the 5.9 GHz band is the Unlicensed National Information Infrastructure (U-NII) 3 band, which is used by Wi-Fi devices for provisioning wireless local area networks. Additionally, the FCC has released a Notice for Proposed Rulemaking to consider unlicensed Wi-Fi operations in the 6 GHz band~\cite{fcc_6_nprm}, which lies at the upper end of the 5.9 GHz band. Furthermore, one of the two mechanisms considered in the US for DSRC--Wi-Fi coexistence~\cite{website:Channelization} proposes to divide the 5.9 GHz band into two sub-bands. The lower 40 MHz will be used for vehicular non-safety applications, and can, therefore, tolerate a slightly higher level of interference. Spectrum sharing between DSRC and Wi-Fi can be permitted in this sub-band. The upper 30 MHz, on the other hand, will be used exclusively for vehicular safety applications and no spectrum sharing will be permitted in this sub-band. This proposal requires DSRC and Wi-Fi systems to operate in adjacent channels, without any guard band. 

\begin{figure}[htb]
    \centering
    \includegraphics[scale=0.35]{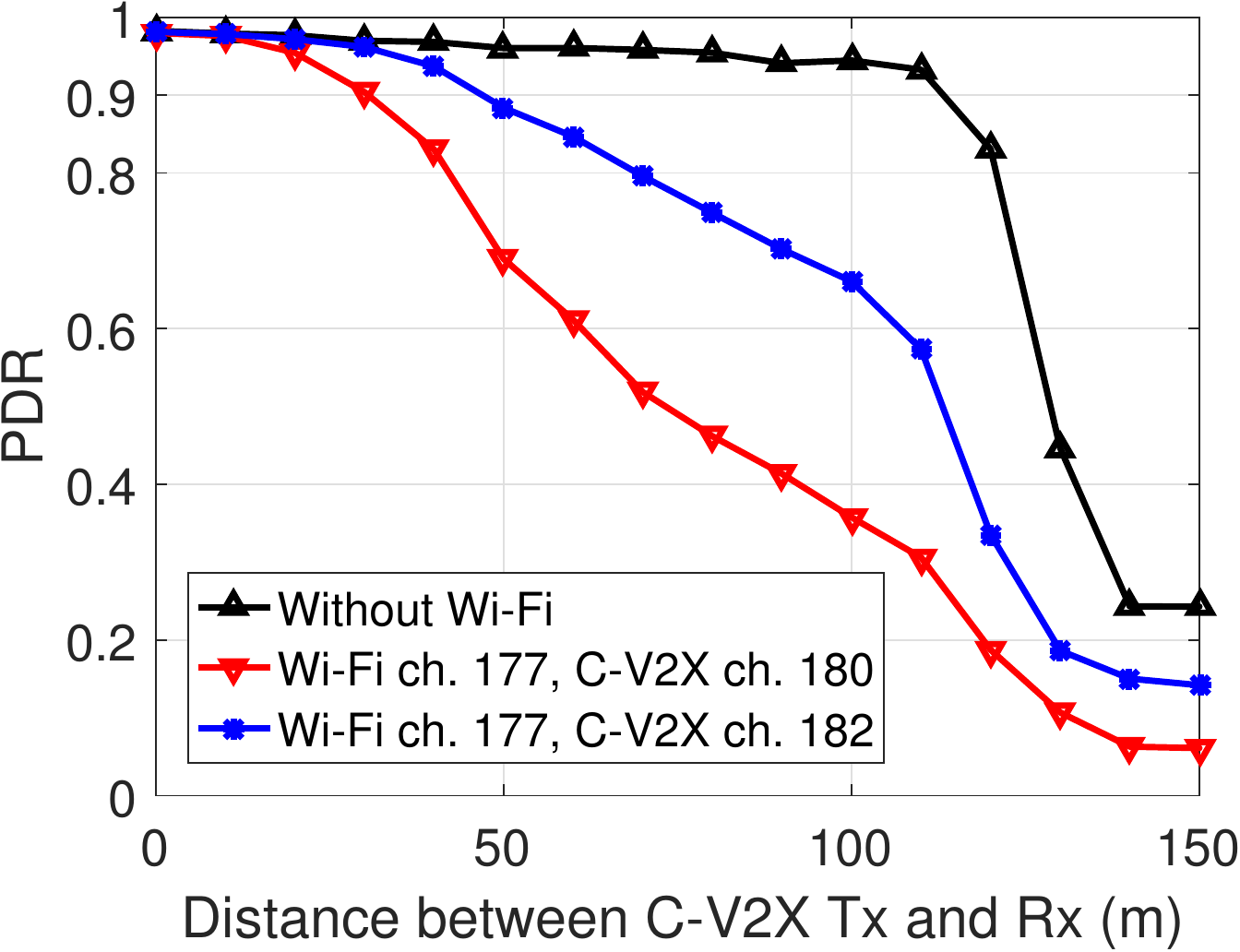}
    \caption {Impact of adjacent channel Wi-Fi on C-V2X mode 4 performance.}
    \label{fig:unii4-system}
    \vspace{-4mm}
\end{figure}

In each of the aforementioned scenarios, regardless of the V2X RAT used, if a Wi-Fi device operating in the adjacent channel is located very close to the V2X receiver, the noise floor of that receiver will be elevated and will inevitably lead to a loss in its performance. The performance loss depends on the frequency separation between the Wi-Fi and the ITS channel. Consider for example a Wi-Fi network operating in channel 177\footnote{The use of channel 177 will be permitted if the U-NII-4 re-channelization proposal~\cite{website:Channelization} is approved.}. Fig.~\ref{fig:unii4-system} shows the impact of adjacent channel Wi-Fi transmissions on the system-wide performance of C-V2X mode 4 (operating in channels 180/182), where the PDR of C-V2X is plotted against the distance between the C-V2X transmitter and receiver. The frequency separation between Wi-Fi and C-V2X is $0$ and $10$~MHz for channels 180 and 182, respectively. The simulated scenario is the Urban Fast scenario~\cite{techreport}, where the Wi-Fi access point is located $10$~m away from the C-V2X receiver at which the performance is observed. C-V2X devices transmit basic safety messages at $10$~Hz, while the Wi-Fi access point has saturated downlink traffic for its ten associated clients. In the simulated scenario, the $90\%$ PDR range is reduced by $85$~m and $65$~m for channels 182 and 180, respectively! Allocations of frequency bands adjacent to the ITS band must take such negative effects of adjacent channel operations into consideration. 

\section{Summary \& Conclusions}

The two evolutionary RATs---802.11bd and NR V2X---are a major overhaul of their respective predecessors and are expected to significantly improve on performance metrics like latency, reliability and the throughput. Some noteworthy features of 802.11bd and NR V2X are outlined in Table~\ref{table:ngv-nrv2x-comparison}. 

\begin{table}[htb]
    \centering
    \caption{Comparison of IEEE 802.11bd \& NR V2X}
    \label{table:ngv-nrv2x-comparison}
    \begin{tabular}{|p{2.2cm}|p{2.0cm}|p{3.3cm}|}
    \hline
        \textbf{Feature} & \textbf{IEEE 802.11bd} & \textbf{NR V2X} \\ \hline
        Base Technology & IEEE 802.11n/ac & 5G NR \\ \hline
        PHY layer & OFDM & SC-FDMA, OFDM\\ \hline
        MAC layer & CSMA & Mode 1: gNodeB scheduling \\ 
         & & Mode 2: Flexible sub-modes \\ \hline
        Interoperability & Yes & Non co-channel \\ \hline
        Backward compatibility  & Co-channel & Not backward compatible \\ \hline
        mmWave support & Yes & Yes \\ \hline
    \end{tabular}
    \vspace{-3mm}
\end{table}

Even though the two RATs share some design objectives, their design principles are largely different. The TGbd is set out to re-design a two decade old technology---802.11p---while including all those enhancements that have made recent Wi-Fi standards, i.e. 802.11n/ac, extremely popular in today's networks. While doing so, care is being taken to retain backward compatibility with 802.11p and to make the new standard suitable for extremely high mobility environments. The backward compatibility requirement for 802.11bd is critical because, at least in a few countries, initial deployments of DSRC-equipped vehicles have already taken place.

On the other hand, building atop 5G NR, NR V2X can leverage many of the newer and advanced PHY and MAC layer techniques and features. NR V2X shows promising signs in terms of relative improvements over its predecessor technology as evident in discussions in Sec.~\ref{sec:ngv-performance} and Sec.~\ref{sec:nrv2x-performance}. One of the reasons for the relatively high gains of NR V2X over C-V2X is that NR V2X starts out with a clean slate, without imposing any backward compatibility constraints on the new RAT despite the fact that C-V2X was standardized only in 2017. However, lack of backward compatibility \emph{may not be} as critical for NR V2X as it is for 802.11bd considering the aggressive time-line of 3GPP Rel. 16 standardization and the time-line of adoption of C-V2X among auto-manufacturers.  

Design of two parallel evolutionary RATs presents regional regulators and auto-manufacturers with two options to choose from, based on regional requirements. However, simultaneous adoption of the two evolutionary RATs (or their predecessors) within a geographical region can result in challenging spectrum management issues and operational difficulties. Such spectrum management concerns need to be pro-actively resolved to exploit the maximum benefits of V2X communication capabilities within vehicles.

\bibliographystyle{ieeetr}
\bibliography{file}

\end{document}